\documentstyle[epsfig,preprint,aps,epsf,floats]{revtex}	
\newcommand{\mt}{m_{\tau}}

\newcommand{\gva}{\gamma_{VA}}
\newcommand{\spsi}{\sin\psi}
\newcommand{\cpsi}{\cos\psi}

\newcommand{\cpsiz}{\cos^{2}\psi}
\newcommand{\szpsi}{\sin 2\psi}

\newcommand{\cth}{\cos\beta}
\newcommand{\cthz}{\cos^{2}\beta}

\newcommand{\ke}{{K_{1}}} 
\newcommand{\kz}{{K_{2}}}

\newcommand{\kv}{{K_{4}}}

\newcommand{\keb}{{\overline{K}_{1}}}
\newcommand{\kzb}{{\overline{K}_{2}}}

\begin{document}
\draft
\tighten
%
%
\preprint{\parbox{5cm}{January 1996\\
TTP95-44\ \\
HUTP-95/A050\\
hep-ph/9601275}}
\title{\mbox{}\protect\\[2cm]
  The Scalar Contribution to $\tau \to K \pi \nu_\tau$}
\author{Markus Finkemeier}
\address{Lyman Laboratory of Physics, Harvard University, Cambridge, 
MA 02138, USA}
\author{Erwin Mirkes}
\address{Institut f\"ur Theoretische Teilchenphysik, Universit\"at
Karlsruhe, 76128 Karlsruhe, Germany}
\maketitle
\begin{abstract}
We consider the scalar form factor in $\tau \to K\pi \nu_\tau$
decays. It receives contributions both from the scalar resonance
$K_0^*(1430)$ and from the scalar projection of off-shell vector resonances.
We construct a model for the hadronic current which includes the vector
resonances $K^*(892)$ and $K^*(1410)$ and the scalar resonance
$K_0^*(1430)$. The parameters of the model are fixed 
by matching to the $O(p^4)$ predictions of chiral perturbation theory. 
Suitable angular correlations of the $K\pi$ system
allow for a model independent separation of the vector and scalar
form factor.  Numerical results for the relevant 
structure functions are presented.
\end{abstract}
%
%
%
%
%
\narrowtext
\everymath={\displaystyle}
\newpage
\section{Introduction}
Semileptonic decays of the $\tau$ lepton into $K^- \pi^0 \nu_\tau$
and $\overline{K^0} \pi^- \nu_\tau$ final states provide
a powerful probe of the strange sector of the weak charged
current. 
Recently, the CLEO and LEP experiments have measured
the branching fractions into $K\pi$ final states with fairly good 
precision \cite{cleo1,cleo2,lep}.
Additional and more detailed information on the underlying
strong interaction physics (the meson dominance model and resonance parameters,
chiral perturbation theory, SU(3) flavor symmetry and isospin symmetry)
can be extracted by an analysis of the hadronic system,
{\it e.g.} by analyzing the invariant mass distribution
and in particular angular distributions of the hadronic system as
described below.

In this paper we will discuss 
the scalar form factor in these $\tau$ decay modes.
The hadronic matrix  elements can be parametrized in terms
of two form factors, the vector form factor $F(Q^2)$ and the
scalar form factor $F_S(Q^2)$. 
Both decay modes are expected to be dominated by the
$K^*(892)$ vector resonance $(J^P=1^-)$,
and in fact in our previous paper \cite{FM95} we 
neglected the scalar form factor.
A scalar resonance ($J^P = 0^+$), however, can
enhance the scalar form factor. A good candidate is the $K_0^*(1430)$,
and indeed there might be experimental evidence for an enhancement
in the $K\pi$ invariant mass spectrum around 
$\sqrt{Q^2} = 1400\,\mbox{MeV}$ \cite{personal}.
The situation, however, is more complicated in two ways.
Firstly, in this energy range, 
there is also  a  vector resonance $K^*(1410)$.
Secondly, off-shell the $K^*(892)$ can have a scalar component, which 
contributes to the scalar form factor. 
Vector and scalar 
contributions can be disentangled experimentally in
 a model independent way
by studying suitable  angular distributions of the $K\pi$ system
\cite{KM1}. The relevant information in encoded in four structure
functions. Three of these structure functions ($W_B, W_{SA}, W_{SF}$)
can be measured even if 
the $\tau$ rest frame cannot be reconstructed. 
Of particular importance are the structure functions
$W_{SA}$ and $W_{SF}$, which 
arise from a nonvanishing spin-zero part of the hadronic
matrix element.
$W_{SA}$ is essentially the spin-zero spectral function 
whereas
$W_{SF}$  arises from the interference of the
spin-one part 
and the spin-zero part 
of the hadronic matrix element.
We will show that a measurement of these structure functions
allows
to establish the  $K_0^*(1430)$ contribution in a model independent way
and might also yield interesting information about the
off-shell behaviour of the $K^*$ propagator.

The paper is organized as follows:
In Sec.~2, we discuss the general structure of the matrix element in terms
of the vector and the scalar form factors. We discuss predictions
from chiral perturbation theory for the low energy limit of the
form factors  and derive a model for the form factors
based on a meson dominance model.
The parameters of the  model are then fixed
by  matching it to the chiral perturbation theory limit.
In Sec.~3, we define the structure functions, which determine the
angular distributions and the decay rate.
Numerical results and our conclusions are finally presented in
Sec.~4 and 5, respectively.

\section{The Hadronic Matrix Elements}

The matrix  element ${\cal{M}}$ for the $\tau$--decay into a kaon and a pion
($m_K=0.493$ GeV, $m_{\pi}=0.139$ GeV)
\begin{equation}
\tau(l,s)\rightarrow\nu(l^{\prime},s^{\prime})
+K(q_{1},m_{K})+\pi(q_{2},m_{\pi}) \label{matrix}
\end{equation}
can be written as a product of a leptonic ($M_{\mu}$) and 
a hadronic vector current ($J^{\mu}$) as
\begin{equation}
{\cal{M}}=\sin\theta_{c}\frac{G}{\sqrt{2}}\, M_{\mu}J^{\mu}
\label{mdef2h}
\end{equation}
In Eq.~(\ref{mdef2h}), 
$G$ denotes the Fermi-coupling constant and  $\theta_c$ is the 
Cabibbo angle.
The leptonic  and hadronic  currents are given by 
\begin{equation}
M_{\mu}=\bar{u}(l^{\prime},s^{\prime})\gamma_{\mu}(g_{V}-g_{A}\gamma_{5})u(l,s)
\label{mdef}
\end{equation}
and
\begin{equation}
J^{\mu}=
\langle K(q_{1})\pi(q_{2})
|V^{\mu}(0)|0\rangle
\label{jdef}
\end{equation}
In Eq.~(\ref{mdef}) $s$ denotes the polarization 4-vector of the $\tau$ lepton
satisfying 
\begin{equation}
l_{\mu}s^{\mu}=0\,\hspace{5mm}
s_{\mu}s^{\mu}= -P^{2}
\label{polvec}
\end{equation}
$P$ denotes the polarization  of the $\tau$ in the laboratory frame.
In the case of $Z$ decays into $\tau$ pairs and
after averaging over the $\tau$ directions the polarization is given by
$P=\frac{-2\,v_{\tau}a_{\tau}}{v_{\tau}^{2}+a_{\tau}^{2}}$ with
$v_{\tau}=-1+4\sin^{2}\theta_{W}$ and $a_{\tau}=-1$;
for lower energies effectively $P=0$.

The most general ansatz for the hadronic
vector current in Eq.~(\ref{jdef}) 
is characterized by two form factors $F^{K\pi}(Q^2)$ and $F^{K\pi}_S(Q^2)$
%
%
\begin{equation}
J^{\mu}=
F^{K\pi}(Q^2) \,\,
\left(g^{\mu \nu} - \frac{Q^\mu Q^\nu}{Q^2}\right) \,\,(q_1 - q_2)_\nu 
\,\, + \,\, F^{K\pi}_S(Q^2) \,Q^\mu
\label{fdef}
\end{equation}
where
\begin{equation}
   Q^\mu = q_1^\mu + q_2^\mu
\end{equation}
The vector form factor $F^{K\pi}(Q^2)$ corresponds to the $J^P = 1^-$ component
of the strange weak charged current, and the scalar form factor 
$F^{K\pi}_S(Q^2)$
to the $J^P = 0^+$ component. 
In the limit of $SU(3)$ flavor symmetry, $m_K^2 = m_\pi^2$,
the vector current is conserved, and $F_S(Q^2)$  vanishes.

In the isospin symmetry limit, the form factor $F$ 
for the two decay modes
$\tau \to K^- \pi^0 \nu_\tau$ and $\tau \to \overline{K^0} \pi^- \nu_\tau$
are related by a simple Clebsch-Gordan factor
\begin{equation}
F^{K^- \pi^0} = 
\frac{1}{\sqrt{2}} F^{\overline{K^0} \pi^-};\hspace{1cm}
F^{K^- \pi^0}_S= 
\frac{1}{\sqrt{2}} F^{\overline{K^0} \pi^-}_S
\end{equation}
We will specify the form factor
$F^{\overline{K^0} \pi^-}$ and $F^{\overline{K^0} \pi^-}_S$
in the remaining part of this section.\\[2mm]

Let us add two brief comments regarding the closely related
decay modes $\tau\rightarrow 2\pi\nu_{\tau}$ and
$\tau\rightarrow 2K\nu_{\tau}$.
First, the corresponding scalar form factors in these decay modes
vanish in the isospin symmetry limit,
because in this limit the vector current is exactly conserved (CVC).
However, CVC could be violated 
due to isospin violation. 
Such a violation could be tested by measuring
the structure functions $W_{SA}$ and $W_{SF}$ (see below). In particular
the structure function $W_{SF}$ is expected to provide
the most sensitive test of CVC due to the interference
of a possible   (small) spin-zero part with the (large) vector part. 
Any nonvanishing contribution to $W_{SA}$ or $W_{SF}$ would be a clear signal
of CVC violation.
Second, note that the scalar sector in these tau decays
is also
sensitive to possible new physics effects such as effects from charged Higgs
exchange and possible CP violation \cite{argonne}.
These latter effects could in principle  also be present in the 
$K\pi$ decay mode.
However, we assume that these effects are
much smaller than the contributions discussed in this paper
and we will neglect them in the following.

\subsection{Chiral perturbation theory}
In this subsection we  discuss the dynamical constraints on the form factors
which can be deduced from chiral Lagrangians.
For very small momentum transfers (small compared with the masses of the 
hadronic resonances), chiral perturbation theory provides  standard model
predictions for the hadronic matrix elements \cite{CFU96}.
The form factors for 
$K^+\rightarrow \pi^0 e^+\nu_e$ and
$K^0\rightarrow \pi^- e^+\nu_e$, which
have been calculated in chiral perturbation theory to one loop
in \cite{cpt},
are related by analytic continuation to those relevant for
$\tau\to K \pi \nu_\tau$ (see also  \cite{truong}).
Expanding the results in a Taylor series around $Q^2 = 0$, one obtains
\begin{eqnarray} \label{eqncpt}
F^{\overline{K^0} \pi^-}(Q^2) & = & f(0) 
   \Big[ 1 + \frac{1}{6} \langle r^2 \rangle_V^{K \pi} Q^2 + \cdots \Big]
\nonumber \\
\nonumber \\
F^{\overline{K^0} \pi^-}_S(Q^2) & = & 
\frac{m_K^2 - m_\pi^2}{Q^2}  f_S(0) 
   \Big[ 1 + \frac{1}{6} \langle r^2 \rangle_S^{K \pi} Q^2 + \cdots \Big]
\end{eqnarray}

At leading order (tree level), chiral perturbation theory predicts
$f(0) = f_S(0) = 1$ and $ \langle r^2 \rangle_V^{K \pi} =
 \langle r^2 \rangle_S^{K \pi} = 0$. 
At next-to-leading order, both $f(0)$ and $f_S(0)$ differ slightly from 
unity.
Numerically, $f(0) = f_S(0) = 0.978$ in the isospin symmetry limit
\cite{cpt,truong}. 
For our purposes, however, this small deviation from unity may be neglected,
because we neglect other effects of the same order of magnitude anyway
(Note that isospin breakage, for example, yields $f^{K^- \pi^0}(0) = 0.988$
and  $f^{\bar{K}^0 \pi-}(0) = 0.977$). Therefore we will assume
\begin{equation}
   f(0) = f_S(0) = 1
\end{equation}

At one-loop order, the chiral perturbation theory predictions for the vector
and scalar radii as determined in \cite{cpt} are numerically given by

\begin{eqnarray}
   \langle r^2 \rangle_V^{K \pi}  & = & 0.37 \, \mbox{fm}^2
   = 9.55 \, \mbox{GeV}^{-2}
   \\[1mm]
   \langle r^2 \rangle_S^{K \pi}  & = & 0.20 \, \mbox{fm}^2
   = 5.14 \, \mbox{GeV}^{-2}
\end{eqnarray}
\subsection{Meson Dominance Model}
The form factors 
in the region of large $Q^2$ are now discussed in the framework of
a resonance exchange model.
We assume that the hadronic current for $\tau \to K \pi \nu_\tau$ is dominated
by three interfering resonances, the vector mesons $K^*(892)$ and 
$K^*(1410) \equiv {K^*}'$
(with $\Gamma_{K^*}=0.05$ GeV, 
$\Gamma_{K^{*'}}=0.227$ GeV)
and the scalar resonance $K_0^*(1430)$
(with $\Gamma_{K_0^{*}}=0.287$ GeV). 
Assuming constant meson couplings and
choosing a certain 
form for the meson propagators
(see the discussion after (\ref{eqnprop})), 
this gives the hadronic current
\begin{eqnarray} 
J^{\overline{K^0} \pi^-\,\,\mu}(q_1,q_2) 
& = & 
   \frac{c_V}  {1 + \beta_{K^*}}
   \left[ \left( g^{\mu\nu} - 
   \frac{Q^\mu Q^\nu}{ m_{K^*}^2} \right) 
   \mbox{BW}_{K^*}(Q^2)\,\, (q_1 - q_2)_\nu \right.
\nonumber \\[2mm]
 & & \qquad { }+ 
\left.    \beta_{K^*}
    \left( g^{\mu\nu} - 
   \frac{Q^\mu Q^\nu}{ m_{{{K^*}'}}^2} \right) 
   \mbox{BW}_{{K^*}'}(Q^2) \,\,(q_1 - q_2)_\nu \right]
\label{eqnmdm} \\[2mm] 
& &\qquad { }
   + \frac{m_K^2 - m_\pi^2}{m_{K_0^*}^2} \,\,c_S \,\,
  \mbox{BW}_{K_0^*}(Q^2)\,\, Q^\mu \nonumber
\end{eqnarray}
Here, $\beta_{K^*}$ describes the relative contribution of the ${K^*}'$ as
compared to the dominant $K^*$. $c_V$ determines the overall strength of the 
vector resonances and contains the relevant meson coupling constants.
Similarly,
$c_S$ determines the strength of the scalar resonance channel. 
In the $SU(3)$ flavour symmetry limit ($m_K=m_{\pi}$), 
this scalar contribution vanishes. 
Therefore
we have factored out a coefficient proportional to the symmetry breaking,
such that $c_S$ is a number of order one.

The Breit-Wigner factors are defined by
\begin{equation}
\mbox{BW}_{X}[Q^2] = 
 {M_{X}^2 \over [M^2_{X}-Q^2-i\sqrt{Q^2} \Gamma_{X}(Q^2)]}\>,
\end{equation}
with energy dependent widths $\Gamma_{X}(Q^2)$ 
\begin{eqnarray}
   \Gamma_{X}(Q^2) & = & \Gamma_X \frac{M_X^2}{Q^2} 
   \left( \frac{p}{p_X} \right)^{2n + 1}
\nonumber \\[2mm]
   p & = & \frac{1}{2 \sqrt{Q^2}} 
   \sqrt{[Q^2 - (M_K^2 + M_\pi^2)^2]\, [Q^2 - (M_K^2 - M_\pi^2)]}
   \\[2mm]
   p_X & = & \frac{1}{2 M_X}
   \sqrt{[M_X^2 - (M_K^2 + M_\pi^2)^2]\, [M_X^2 - (M_K^2 - M_\pi^2)]}
\nonumber
\end{eqnarray}
where $n=1$ for the 
$K^*$ and the ${K^*}'$ (p-wave phase space) and $n=0$ 
for the  $K_0^*$ (s-wave phase space)  \cite{Dec93}.

Comparing Eq.~(\ref{eqnmdm}) with Eq.~(\ref{fdef}) yields
\begin{eqnarray} 
   F^{\overline{K^0} \pi^-}(Q^2) & = & \frac{c_V}{1 + \beta_{K^\star}}
  \Big[ \mbox{BW}_{K^\star}(Q^2) + \beta_{K^\star}\, 
   \mbox{BW}_{{K^\star}'}(Q^2)
  \Big]
\label{fkpi}
\\[2mm]
F_S^{\overline{K^0} \pi^-}(Q^2) & = & 
      \frac{m_K^2 - m_\pi^2}{Q^2} \frac{c_V}{1 + \beta_{K^*}}
   \left[ \frac{m_{K^*}^2 - Q^2}{m_{K^*}^2} \mbox{BW}_{K^*}(Q^2) 
 \, + \, \beta_{K^*} \frac{m_{{K^*}'}^2 - Q^2}{m_{{K^*}'}^2} 
   \mbox{BW}_{{K^*}'}(Q^2)    \right]
\nonumber \\[2mm] & & 
+ \,\frac{m_K^2 - m_\pi^2}{m_{K_0^*}^2} \,\,c_S \,\, \mbox{BW}_{K_0^*}(Q^2)
\label{eqnff}        \\[2mm] 
  & \approx & 
  (m_K^2 - m_\pi^2) \Big[    \frac{c_V}{Q^2} 
\,+ \,\frac{c_S}{m_{K_0^*}^2}  \mbox{BW}_{K_0^*}(Q^2)
  \Big] \nonumber
\end{eqnarray}
where the approximation in the last line
\begin{equation}
   \frac{m_{K^*}^2 - Q^2}{m_{K^*}^2} \mbox{BW}_{K^*}(Q^2)
   \approx 
   \frac{m_{K^*}^2 - Q^2}{m_{K^*}^2} \frac{m_{K^*}^2}{m_{K^*}^2 - Q^2}
   = 1
\end{equation}
(and similarly for $K^*\rightarrow K^{*'}$)
becomes exact in the narrow width limit $\Gamma_{K^*}$, 
$\Gamma_{{K^*}'} \to 0$, or below threshold, $Q^2 < (m_K + m_\pi)^2$.
Note that this limit implies that the off-shell 
contribution of the vector resonances
to the scalar form factor is effectively non-resonant. The Breit-Wigner
enhancement is canceled by the off-shellness requirement. 
Thus the $K_0^*$ will be dominant in $F_S$
unless $c_S$ is unnaturally small. 
\subsection{Matching}
We now  match the meson dominance model to chiral perturbation
theory .
Expanding the form factor predictions
from the 
 meson dominance model in Eqs.~(\ref{fkpi},\ref{eqnff}) around $Q^2 = 0$ yields
\begin{eqnarray} 
   F^{\overline{K^0} \pi^-}(Q^2) & = & c_V
   \left[ 1 + \frac{1}{1 + \beta_{K^*}} \left(
   \frac{1}{m_{K^*}^2} + \frac{\beta_{K^*}}{m_{{K^*}'}^2} \right)
   Q^2 + \cdots \right]
 \\[2mm]
  F_S^{\overline{K^0} \pi^-}(Q^2) & = & 
      \frac{m_K^2 - m_\pi^2}{Q^2} 
    \left[ c_V + \frac{c_S}{m_{K_0^*}^2} Q^2 + \cdots \right]
\end{eqnarray}
We now compare this to the constraints from chiral perturbation
theory in Eq.~(\ref{eqncpt}). Matching the leading order terms yields
\begin{equation}
   c_V =  1
\end{equation}
Note that with this  choice, we simultaneously reproduce the correct
leading terms for the 
vector and scalar form factors. This relies on our choice for the meson 
propagators in Eq.~(\ref{eqnmdm}).
Due to the fact that the mesons are objects with structure, the
form of their propagators becomes model dependent in the off-shell
region \cite{Isg}.
In Eq.~(\ref{eqnmdm}), we 
used a $K^*$ propagator proportional to
\begin{equation} \label{eqnprop}
      \left( g^{\mu\nu} -  \frac{Q^\mu Q^\nu}{ m_{K^*}^2} \right)
\end{equation}
and not, as sometimes suggested, proportional to
\begin{equation} \label{eqnprop1}
     \left( g^{\mu\nu} -  \frac{Q^\mu Q^\nu}{ Q^2} \right)
\end{equation}
instead.
The latter form 
would have led to a vanishing contribution 
of the vector resonances to the scalar form factor, 
and we would not
be able to reproduce the correct low energy limit. 
Note that the scalar resonance does not enter at this point,
because its contribution starts at higher order in
${\cal O}(Q^2)$.
With the form of Eq.~(\ref{eqnprop}), however, we simultaneously reproduce
the correct low energy limit for the vector and the scalar form
factor in a natural way.
This gives some credibility to our choice for the off-shell 
propagator and the vector meson dominance model.

Matching the next-to-leading terms requires
\begin{eqnarray} 
   \frac{1}{1 + \beta_{K^*}} \left( 
   \frac{1}{m_{K^*}^2} + \frac{\beta_{K^*}}{m_{{K^*}'}^2} \right)
   & = & \frac{\langle r^2 \rangle_V^{K \pi}}{6}
\label{eqnmatch}   \\[2mm]
   \frac{c_S}{m_{K_0^*}^2} & = & \frac{\langle r^2 \rangle_S^{K \pi}}{6}
\end{eqnarray}
The second equation determines $c_S$ to 
\begin{equation} \label{eqncs}
   c_S  =  \frac{m_{K_0^*}^2}{6} \langle r^2 \rangle_S^{K \pi} = 1.7
\end{equation}
We would like to emphasize 
that the exact numerical value for $c_S$ should not
be taken too seriously, in view of the large extrapolation from $Q^2 = 0$ 
to $Q^2 = m_{K_0^*}^2$.

Eq.~(\ref{eqnmatch}) could be used to determine
$\beta_{K^*}$ (resulting in $\beta_{K^*} \approx - 0.3$).
Note, however, that a small uncertainty in $\langle r^2 \rangle_V^{K \pi}$
translates into a large uncertainty in $\beta_{K^*}$.
Therefore we will instead use the value for $\beta_{K^*}  =-0.135$,
which has been determined in \cite{FM95} from a fit to the
branching ratio%
\footnote{
In \cite{FM95}, we used the value ${\cal B}(K \pi\nu_\tau) 
= 1.36 \pm 0.08 \%$ from \cite{Hel94}.
This yielded $\beta_{K^*} = - 0.135$, very close to the corresponding
strength of the $\rho'$ contribution to the $\rho$ in $\tau\to \pi\pi
\nu_\tau$ \cite{KueSa}, as is expected in the limit of $SU(3)$ flavor
symmetry.
In the meanwhile, a new preprint by CLEO has appeared \cite{cleo1},
with new data on the $\tau\to\bar{K}^0 \pi^- \nu_\tau$. 
Adding to this mode the CLEO 94 data \cite{cleo2}
on $\tau\to K^- \pi^0\nu_\tau$
yields ${\cal B}(K \pi\nu_\tau) = 1.21 \pm 0.14 \%$.
However, additionally taking the isospin constraint into account,
(the $2:1$ ratio between the two modes), which is not well 
satisfied by the data, CLEO obtains
${\cal B}(K \pi\nu_\tau) = 1.11 \pm 0.12 \%$.
Note that this fit value is several standard deviations
lower than the old world average.
It leads to a value of $\beta_{K^*}$ of about
$\beta_{K^*}  =-0.05 \pm 0.05$, compatible with zero.
In the rest of this paper we will use $\beta_{K^*} = - 0.135$ as 
standard value, but we also compare with $\beta_{K^*} = 0$.
}
of $\tau\to K \pi \nu_\tau$. 
With this value,
the left hand  and the right hand sides of the first equation in 
Eq.~(\ref{eqnmatch}) are $1.37\,\mbox{GeV}^{-2}$ and $1.59\,\mbox{GeV}^{-2}$,
respectively. 
In view of the uncertainty of $\langle r^2 \rangle_V^{K \pi}$, 
mainly from higher orders in chiral perturbation theory, 
this $15 \%$ mismatch is certainly acceptable.

\section{Decay Rate, Angular Distributions and Structure Functions}
In this section, we introduce the formalism of the structure
functions which allow for a model independent separation 
of the vector and scalar contribution to the $K\pi$ mode.
Interesting additional informations, such as the relative
sign of spin-zero and spin-one part in the hadronic
matrix element, will also become possible through the measurement
of the structure functions.
The formalism is largely based on 
\cite{KM1}. We will, however, specify the results in \cite{KM1} for 
the tau decay mode into $K\pi\nu_\tau$.

The differential decay rate 
is obtained from
\begin{equation}
d\Gamma(\tau\rightarrow \nu_{\tau}K\pi)=\frac{1}{2\mt} 
\frac{G^{2}}{2}\sin^{2}\theta_{c}\,
\left\{L_{\mu\nu}H^{\mu\nu}\right\}
\,d\mbox{PS}^{(3)}
\label{decay}
\end{equation}
where  $L_{\mu\nu}=M_{\mu}(M_{\nu})^{\dagger}$ and
$H^{\mu\nu}=J^{\mu}(J^{\nu})^{\dagger}$ with $M_\mu$ and
$J^\mu$ defined in Eqs.~(\ref{mdef},\ref{jdef}).

In the following, the $\tau\rightarrow  K\pi \nu_{\tau}$
decay mode is  analyzed in the hadronic rest frame
$\vec{q}_{1}+\vec{q}_{2}=0$.
After integration over the unobserved
neutrino direction, the phase space
is parametrized in the $K$-$\pi$-rest frame  by
\begin{eqnarray}
\,d\mbox{PS}^{(3)} &=&
\frac{1}{(4\pi)^{3}}
\frac{(\mt^{2}-Q^{2})}{\mt^{2}}\,
               |\vec{q}_{1}|\,
     \frac{dQ^{2}}{\sqrt{Q^{2}}}  \,
      \frac{d\cos\beta}{2}\,  \frac{d\cos\theta}{2}\label{gamma}
\end{eqnarray}
The angle $\beta$ in Eq.~(\ref{gamma})
denotes the angle between the direction
of  the kaon ($\hat{q}_{1}=\vec{q}_{1}/|\vec{q}_{1}|$) 
and the direction of the laboratory $\vec{n}_{L}$ viewed
from the hadronic rest frame
\begin{equation}
\cos\beta = \vec{n}_{L}\cdot \hat{q}_{1}
\end{equation}
$\vec{n}_{L}$  is obtained from $\vec{n}_{L}=-\vec{n}_{Q}$, where
$\vec{n}_{Q}$ denotes the direction of the $K$-$\pi$--system in the
laboratory\footnote{Since the $\tau$ direction $\vec{n}_{\tau}$ as seen from
the hadronic rest frame cannot in general be determined in the 
present $e^{+}e^{-}$--experiments, it is not possible to study 
the angular distribution of $\hat{q}_{1}$ with respect to $\vec{n}_{\tau}$.}.

The angle $\theta$  ($0\leq\theta\leq\pi$) 
in Eq.~(\ref{gamma})  is the angle between the
direction of flight of the $\tau$  in the laboratory frame (= direction of the $\tau$
polarization)
and the direction of the  hadrons as seen in the $\tau$ rest frame.
Note that the cosine of the angle $\theta$
 is observable even in experiments, where
the $\tau$ 
direction cannot be measured experimentally.
This is because $\cos\theta$ 
can be  calculated \cite{KW84,KM2,KM1}
from the energy $E_{h}$ of the hadronic system
with respect to the laboratory frame
\begin{equation}
\cos\theta = \frac{\left(2x\mt^{2}-\mt^{2}-Q^{2}\right)}{
          (\mt^{2}-Q^{2}) \sqrt{1-4\mt^{2}/s}}
\label{cthdef}
\end{equation}
with
\begin{equation}
\hspace{0.1cm}  x  = 2\frac{E_{h}}{\sqrt{s}}
\hspace{1cm}    s = 4 E^{2}_{{beam}}
\end{equation}
Of particular importance  for the subsequent discussion is $\psi$, the angle 
between the direction
of  the laboratory and the $\tau$ as seen from the hadronic rest frame.
Again the cosine of this angle can be
calculated from the hadronic energy $E_{h}$ \cite{KW84,KM2,KM1}.
One has:
\begin{eqnarray}
\cos\psi &=&
       \frac{x(\mt^{2}+Q^{2})-2Q^{2}}{(\mt^{2}-Q^{2})\sqrt{x^{2}-4Q^{2}/s}}
\end{eqnarray}
We are now in the position to evaluate the lepton and hadron tensors
$L_{\mu\nu}$ and $H^{\mu\nu}$ in Eq.~(\ref{decay}).

The angular $(\beta)$ and $E_{h}$ (through $\theta$ and $\psi$) dependence
of the matrix element $L_{\mu\nu}H^{\mu\nu}$ can be disentangled by introducing
suitable linear combinations of density matrix elements of the
hadronic-system\footnote{
The most general decomposition of $L_{\mu\nu}H^{\mu\nu}$ (for two body decays)
in terms
of density matrix elements (or structure functions) $W_X$
of the hadronic system has two additional terms
$\bar{L}_{A}W_{A}+\bar{L}_{E}W_{E}$ \cite{KM1}. However,
$W_A$ and $W_E$ vanish in the case of tau decays into two
pseudoscalar mesons. Nonvanishing $W_A$ and $W_E$ arise for example in
decay modes with a vector and a pseudoscalar \cite{dm1}.)
}:
\begin{equation}
L_{\mu\nu}H^{\mu\nu}=
(g_{V}^{2}+g_{A}^{2})(\mt^{2}-Q^{2})\,
(\,\bar{L}_{B}W_{B}+\bar{L}_{SA}W_{SA}+\bar{L}_{SF}W_{SF}
  +\bar{L}_{SG}W_{SG}\,)
\label{lh}
\end{equation}
The decomposition of the lepton- and hadron-tensor in Eq.~(\ref{lh})
has the advantage that the 
coefficients $L_{B,SA,SF,SG}$ contain
all $\beta,E_{h}(\theta,\psi)$  and 
$\tau$--polarization dependence (see below), whereas the hadronic structure
functions  
$W_{B,SA,SF,SG}$ depend only on $Q^2$ and the form factors $F$ and $F_S$
of the hadronic current.
The angular coefficients $\bar{L}_{B,SA,SF,SG}$ 
in Eq.~(\ref{lh}) are given by \cite{KM1}:
\begin{eqnarray}
\bar{L}_{B} &=&     {2}/{3}\,\ke\,+\,\kz
         \,-\,  2/3\,\keb\,(3\cthz-1)/2 \nonumber  \\
\bar{L}_{SA} &=&    \kz   \label{ldef}\\
\bar{L}_{SF}       &=&  - \kzb\,\cth \nonumber\\
\bar{L}_{SG}       &=&  0 \nonumber
\end{eqnarray}
with
\begin{eqnarray}
\ke &=&  1-\gva P \cos\theta -
        ({\mt^{2}}/{Q^{2}})\, (1+ \gva P \cos\theta)  \nonumber\\[1mm]
\kz &=&  ({\mt^{2}}/{Q^{2}})\, (1+ \gva P \cos\theta)  \nonumber\\[1mm]
\keb&=&  \ke \,(3\cpsiz-1)/2 - 3/2\,\kv\,\szpsi\nonumber \\[1mm]
\kzb&=& \kz\,\cpsi\,+\,\kv\,\spsi \label{kalldef}\\[1mm]
\kv &=&   \sqrt{{\mt^{2}}/{Q^{2}}}\,\,\gva\, P \sin\theta \nonumber
\end{eqnarray}
where
\begin{equation}
\gamma_{VA}= \frac{2g_{V}g_{A}}{g_{V}^{2}+g_{A}^{2}}
\label{gamvadef}
\end{equation}
In the standard model $\gamma_{VA}=1$.\\
For $P=0$, the case relevant in the low energy
region $\sqrt{S}\approx 10$ GeV,  Eq.~(\ref{ldef}) simplifies to
\begin{eqnarray}
\bar{L}_{B}  &=&  \frac{1}{3}\left(2+\frac{\mt^2}{Q^2}\right)
            \, - \, \frac{1}{6}\left(1-\frac{\mt^2}{Q^2}\right)
                    \left(3\cpsiz-1\right)
                    \left(3\cos^2\beta-1\right)  \nonumber  \\
\bar{L}_{SA} &=&    \frac{\mt^2}{Q^2}    \\
\bar{L}_{SF} &=&    -\frac{\mt^2}{Q^2}\,\cpsi\,\cos\beta   \nonumber 
\end{eqnarray}
Note that the angular coefficient $\bar{L}_{SG}$ vanishes, if the hadronic
rest frame is experimentally not known and only the distribution 
in $\beta$ is considered.
One may alternatively conceive of an experiment
where the $\tau$ direction can be determined. 
This would allow to analyze the distribution of the hadronic system
in term of two kinematical angles \cite{KM1} 
and in particular, $\bar{L}_{SG}$
would not vanish. The relevant angular coefficients can be found
in appendix B of \cite{KM1}.

The hadronic structure functions  $W_{B,SA,SF,SG}$ 
are related to the form factors in Eq.~(\ref{fdef}).
The dependence can be obtained from
Eq.~(34) in \cite{KM1}, with the replacements
$
x_4  \rightarrow 2 \,\vec{q}_1 \>,
F_3  \rightarrow -iF\, \>,
F_4  \rightarrow   F_S\>.
$
One has:
\begin{eqnarray}
W_B    &=& 4 (\vec{q}_1)^2\,|F|^2\, \nonumber  \\
W_{SA} &=& Q^2\,  |F_S|^2\,     \label{w}      \\
W_{SF} &=& 4\sqrt{Q^2}|\vec{q}_1|  \, \mbox{Re}\left[FF_S^*\right]\nonumber\\
W_{SG} &=& -4\sqrt{Q^2}|\vec{q}_1|  \, \mbox{Im}\left[FF_S^*\right]\nonumber
\end{eqnarray}
where $|\vec{q}_1|=q_1^z$ is the momentum of the kaon 
in the rest frame of the hadronic system.
\begin{equation}
q_{1}^{z}=\frac{1}{2\sqrt{Q^{2}}}\left(
[Q^{2}-m_1^{2}-m_2^{2}]^2-4m_{1}^{2}m_{2}^{2}\right)^{1/2}
\quad E^2_1=(q_1^{z})^2+m^2_1 
\end{equation}

The hadronic structure functions
$W_{B,SA,SF,SG}$ are linearly related to the density matrix elements 
of the $K$-$\pi$ system:
\begin{equation}
\begin{array}{lrccc}
W_{B}  & =& \hspace{3mm} \tilde H^{00}         &=&     H^{33}        \\
W_{SA} & =& \hspace{3mm} \tilde H^{ss}         &=&     H^{00}        \\
W_{SF} & =& \hspace{3mm} 
                   \tilde H^{s0}+\tilde H^{0s} &=&     H^{03}+H^{30} \\
W_{SG} & =& \hspace{3mm}- i
                 (\tilde H^{s0}-\tilde H^{0s}) &=& -i(H^{03}-H^{30}) \\
    \end{array}
\label{hidef}
\end{equation}
where
\begin{equation}
\tilde H^{\sigma\sigma^{\prime}}=\epsilon_{\mu}(\sigma)H^{\mu\nu}
                \epsilon_{\nu}^{\ast}(\sigma^{\prime})
\label{hssdef}
\end{equation}
and
\begin{equation}
\begin{array}{ll}
\epsilon_{\mu}(s)  =(1;0,0,0)\hspace{1cm}&
\epsilon_{\mu}(0)  =(0;0,0,1)\\[2mm]
\end{array}
\label{polvekdef}
\end{equation}
are the polarization vectors
for a hadronic system in a spin one $(\epsilon(0))$ 
or spin zero  $(\epsilon(s))$  state with the $z$--direction 
aligned with $\hat{q}_{1}$.
The r.h.s of Eq.~(\ref{hidef}) refers to the space-space components
$H^{mn}=H_{mn}$ ($m,n=0,1,2,3$) of $H^{\mu\nu}$.

With Eqs.~(\ref{decay}-\ref{polvekdef}) 
the most general angular distribution of the $K\pi$ final state
in the decay of a polarized $\tau$ is presented.
Integrating over the angles $\beta$ and $\theta$ we obtain
the formula for the differential decay rate $d\Gamma/dQ^2$:
\begin{eqnarray} \label{eqndiff}
\frac{d\Gamma}{dQ^2}&=&
           \frac{G^{2}\sin^{2}\theta_{c}}{2^8\pi^3} 
(g_{V}^{2}+g_{A}^{2})
\frac{(\mt^{2}-Q^{2})^{2}}{\mt Q^{3/2}}\,
\,|{q}_{1}^z|\,\,
\frac{2Q^2+\mt^2}{3\mt^2}
 \left\{ W_B \,+ \,\frac{3\mt^2}{2Q^2+\mt^2} \,W_{SA}
 \right\}
\end{eqnarray}
\nopagebreak{
\section{Numerical Results}

\setlength{\unitlength}{0.7mm}
\begin{figure}[hbt]               \vspace*{-2cm}
\begin{picture}(150,165)(-30,1)
\mbox{\epsfxsize10.0cm\epsffile[78 222 480 650]{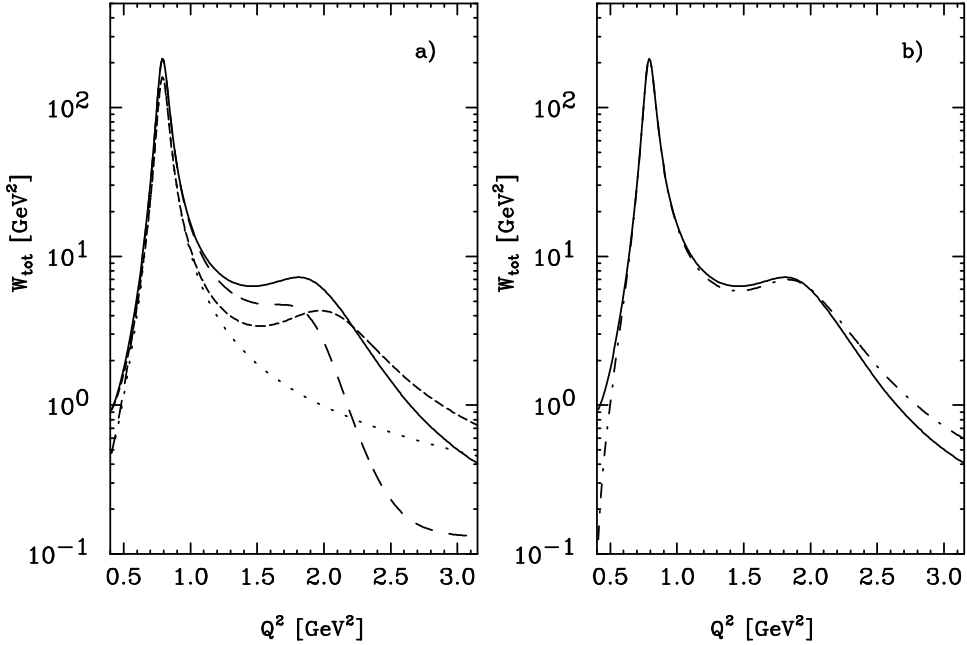}}
\end{picture}
\caption{
a)
$Q^2$ dependence of $W_{tot}$ for
$\beta_{K^*}=-0.135,\, c_S=1.7$ (solid);\protect\\
$\beta_{K^*}=-0.135,\, c_S=0$ (long dashed);
$\beta_{K^*}= 0,\, c_S=1.7$ (short dashed);
$\beta_{K^*}= 0,\, c_S=0$ (dotted).\protect\\[1mm]
b)
The $Q^2$ dependence of $W_{tot}$ for
$\beta_{K^*}=-0.135,\, c_S=1.7$ is compared
with the prediction 
from the purely transverse vector meson propagator 
in (\protect \ref{eqnprop1}) (dotted-dashed)
}
\label{fig1}
\end{figure}
}

\setlength{\unitlength}{0.7mm}
\begin{figure}[hbt]               \vspace*{-2cm}
\begin{picture}(150,165)(-30,1)
\mbox{\epsfxsize10.0cm\epsffile[78 222 480 650]{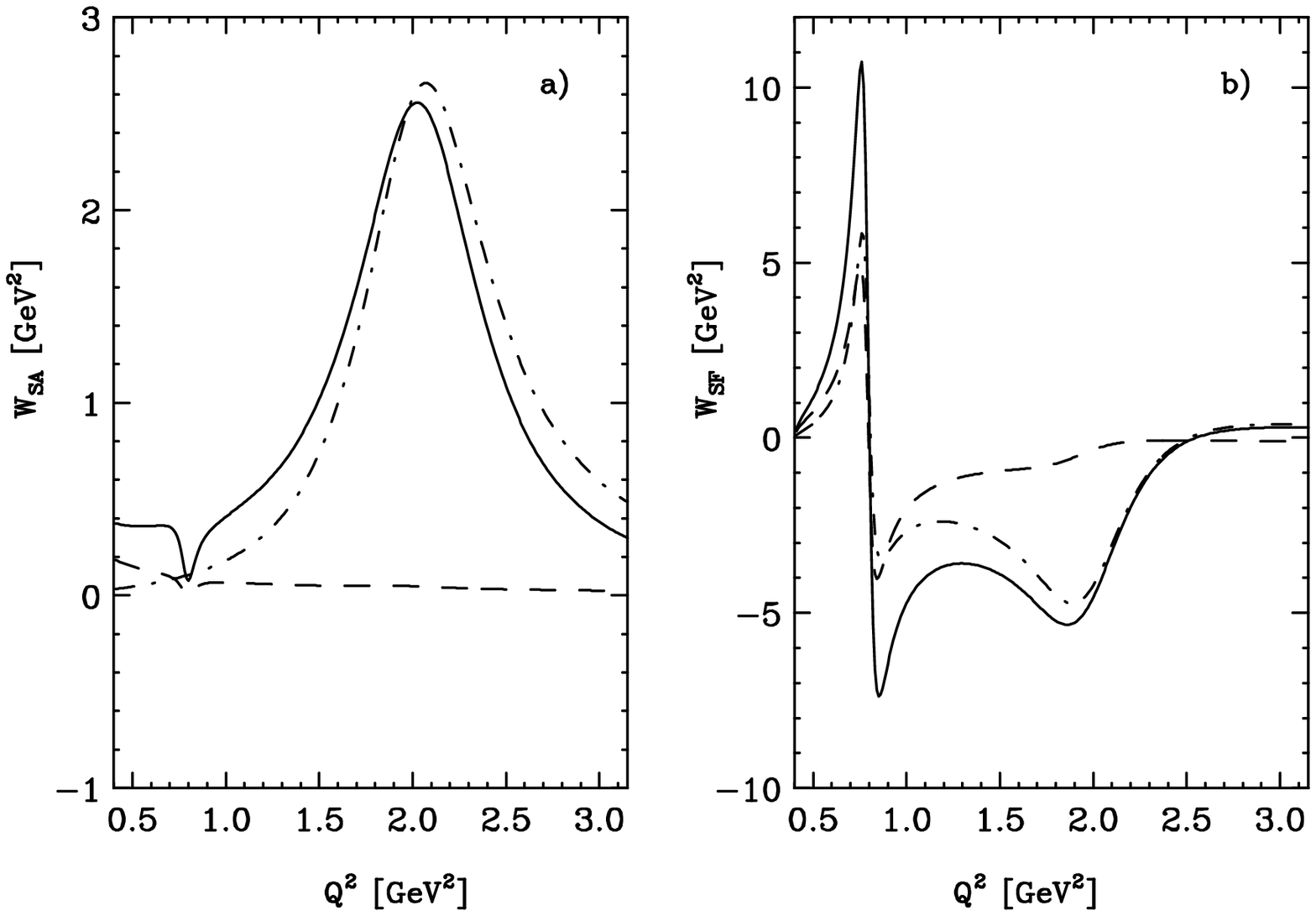}}
\end{picture}
\caption{
$Q^2$ dependence of $W_{SA}$ (a) and $W_{SF}$ (b) for
$ c_S=1.7$ (solid),
$ c_S=0$ (long dashed),
$ c_S=1.7$ but without the  scalar off-shell $K^*$ contributions
(dotted-dashed). The parameter  $\beta_{K^*}$ is fixed to -0.135.
}\label{fig2}
\end{figure}

The following numerical studies are shown for the full $K\pi$ final
state, {\it i.e.} with  a contribution
of one third from $K^- \pi^0$ mode and two thirds from 
$\bar{K}^0 \pi^-$.

In Fig.~\ref{fig1} we display $W_{tot}(Q^2)$, which is defined as
\begin{equation}
  W_{tot}(Q^2) = 
   W_B (Q^2) \,+ \,\frac{3\mt^2}{2Q^2+\mt^2} \,W_{SA}(Q^2)  
\end{equation}
and which according to Eq.~(\ref{eqndiff}) determines the kaon-pion invariant
mass spectrum $d\Gamma/dQ^2$.

The solid line  gives the prediction from our model,
including the $K_0^*$ resonance (with $c_S = 1.7$) and the ${K^*}'$ 
resonance (with $\beta_{K^*}=-0.135$). 
We predict a clearly visible shoulder around
$Q^2 \approx 2\,\mbox{GeV}^2$.
In Fig.~1a we compare with 
the results obtained without the $K_0^*$, {\it i.e.}\ with
$c_S = 0$, and without the ${K^*}'$, i.e.\ with $\beta_{K^*} = 0$. 
The two resonances lead to a sizable 
enhancement  around $Q^2 \approx 2\,\mbox{GeV}^2$.
Within our model,  we find that the $K^{*'}$ contributes 
mainly below its on-shell mass $Q^2 = 2\,
\mbox{GeV}^2$, because the negative interference with the tail of the
$K^*(892)$ is negative above this value. 
Therefore an enhancement in the $Q^2 = 
2.0 \cdots 2.5 \, \mbox{GeV}$ regime would give some indication of
the presence of a $K_0^*$ contribution.
This conclusion, however, is  model dependent because
it depends on  
the relative phase of the $K^*(892)$ and the $K^{*'}$ 
(negative $\beta_{K^*}$).
Predictions for the $Q^2$ dependence of $W_{tot}$ without higher
mass resonances $K^{*'}$ and $K_0^*$ are shown as the dotted curve
in Fig.~1a.

The branching ratios corresponding to the various predictions
in Fig.~1a are (the relative  contribution  from
the scalar current is given in brackets):
$$
\begin{array}{l@{\quad}l@{\quad}l@{\quad}l}
\beta_{K^*}=-0.135, & c_S=1.7: & {\cal B}(K\pi)= 1.41\% & [4\%] \\
\beta_{K^*}=-0.135, & c_S=0.0: & {\cal B}(K\pi)= 1.36\% & [0.6\%]\\
\beta_{K^*}= 0    , & c_S=1.7: & {\cal B}(K\pi)= 1.04\% & [5.7\%]\\
\beta_{K^*}= 0    , & c_S=0  : & {\cal B}(K\pi)= 0.99\% & [0.7\%].
\end{array}
$$
The value of $c_S$ hardly affects the total decay rate.
The effect of $\beta_{K^*}$ is basically due to the change in the
normalization of the $K^*(892)$ contribution in Eq.~(\ref{fkpi}).
The effect of the direct $K^*(1410)$ resonance contribution is 
phase space suppressed and similar in size to the effect of
the $K_0^*$ contribution via $c_S$.

In Fig.~1b, we explore the sensitivity of the invariant mass distribution
to the scalar projection of the off-shell vector resonances
$K^*$ and $K^{*'}$.
The prediction for  $W_{tot}$ from our model with the 
vector propagator in Eq.~(\ref{eqnprop}) is compared with the prediction
based on the purely transverse vector propagator in Eq.~(\ref{eqnprop1}).
Sizable differences are found only very small  ($Q^2<0.6$ GeV$^2$) or
very large ($Q^2 > 2.4$ GeV$^2$) invariant mass, where the rate is
extremely small. 
Note that the difference in $W_{tot}$ in Fig.~1b
is entirely due to the difference in $W_{SA}$.

The studies in Fig.~1 show
that from a measurement of the $K \pi$ invariant
mass spectrum alone,
a scalar resonance contribution from the 
$K_0^*$ or scalar off-shell effects from vector resonances
can not be established reliably.
We will show in the following that
the formalism of the structure functions
allows for a much more detailed and model independent investigation
of these effects. 
Remember that a measurement of $W_{SA}$ separates the
scalar from the vector contribution to the rate, whereas $W_{SF}$ and
$W_{SG}$ measure the real and imaginary part of the interference
term.

\setlength{\unitlength}{0.7mm}
\begin{figure}[ht]               \vspace*{-2cm}
\begin{picture}(150,165)(-30,1)
\mbox{\epsfxsize10.0cm\epsffile[78 222 480 650]{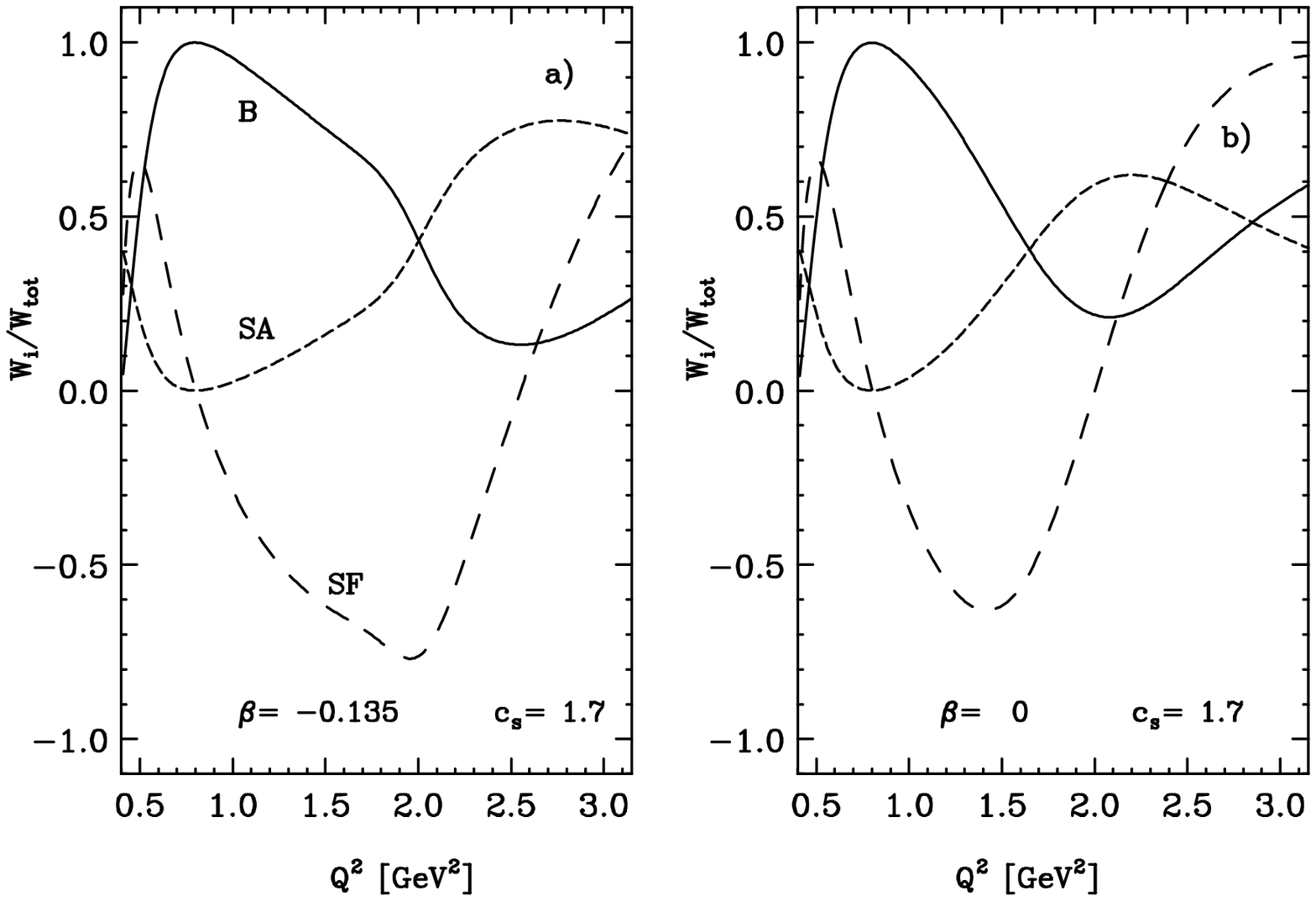}}
\end{picture}
\mbox{}\\[-2cm]
\begin{picture}(150,165)(-30,1)
\mbox{\epsfxsize10.0cm\epsffile[78 222 480 650]{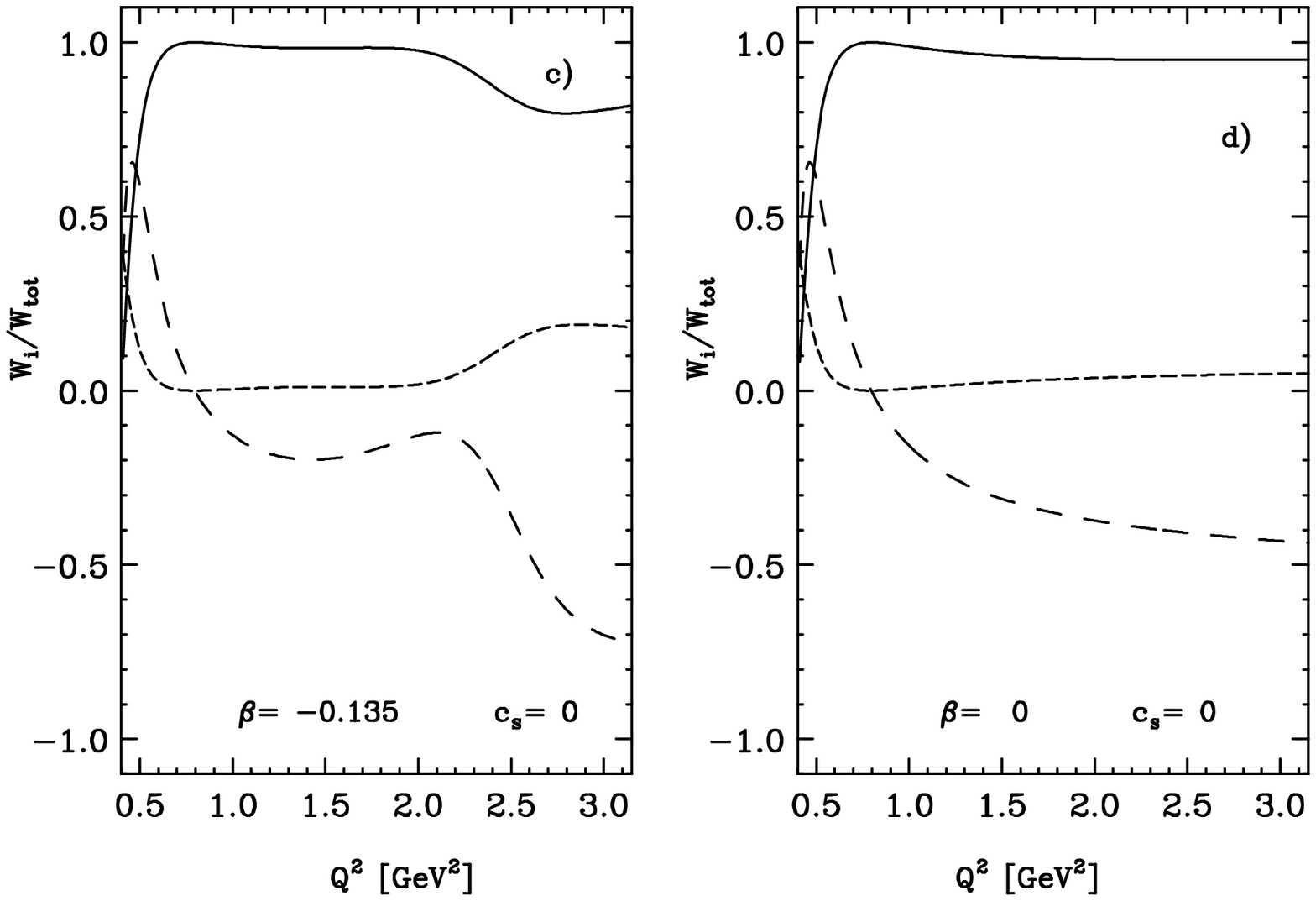}}
\end{picture}
\caption{
Normalized structure functions $W_B/W_{tot}$ (solid),
$W_{SA}/W_{tot}$ (short dashed),
$W_{SF}/W_{tot}$ (long dashed),
for $c_S=1.7$ (a,b) and $c_S=0$ (c,d).
The parameter $\beta_{K^*}$ is $-0.135$ in (a,c) and $0$ in (b,d).
}\label{fig3}
\end{figure}

We give the predictions for the structure function
$W_{SA}$ in Fig.~2a. 
One observes a clear peak from the $K_0^*$ resonance around $2.1 \,
\mbox{GeV}^2$, which is absent for $c_S = 0$. Thus an experimental
confirmation of this enhancement would firmly establish the
$K_0^*$ contribution.
Note that the value of the parameter $\beta_{K^*}$ hardly affects
$W_{SA}$, as is obvious from Eq.~(\ref{eqnff}). 

Additional information about the spin zero part of the matrix
element (in particular the relative sign of the spin zero and the 
spin one part) can be extracted from the structure function
$W_{SF}$, which is shown in Fig.~2b. 
This structure function changes sign at the $K^*$
resonance and receives fairly large contributions due to the interference
effects with the (large) spin one part of the matrix element.
In contrast to $W_{SA}$ this structure function receives  sizable contributions
even without the presence of the scalar resonance $K_0^*$ (dashed curve).
Furthermore,  there is some dependence on the higher $K^{*'}$ contribution
in $W_{SF}$ (see below).

\clearpage


\setlength{\unitlength}{0.7mm}
\begin{figure}[hbt]               \vspace*{-2cm}
\begin{picture}(150,165)(-30,1)
\hspace{3cm}\mbox{\epsfxsize10.0cm\epsffile[78 222 480 650]{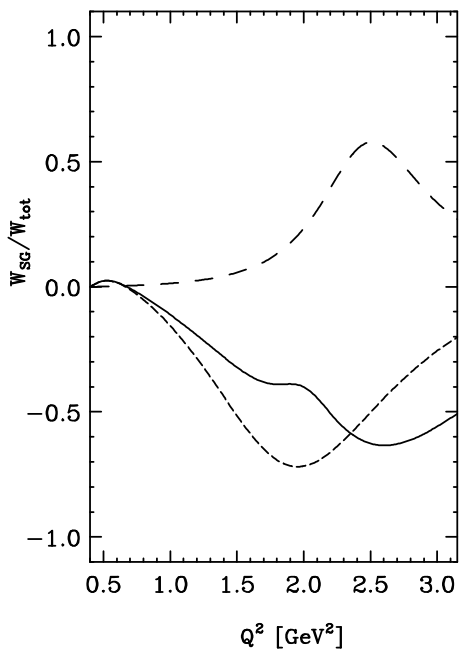}}
\end{picture}
\caption{
Normalized structure function $W_{SG}/W_{tot}$ as a function
of $Q^2$ for
\protect $\beta_{K^\star}=-0.135$, $c_S=1.7$ (solid);
\protect $\beta_{K^\star}=-0.135$, $c_S=0$ (long dashed);
\protect $\beta_{K^\star}= 0$,     $c_S=1.7$ (short dashed).
}\label{fig4}
\end{figure}

In Figs.~2a and b, we also compare the results from the two
different vector meson propagators in 
Eqs.~(\ref{eqnprop},\ref{eqnprop1}).
Regarding $W_{SA}$, we find in Fig.~2a that the interference of the
$K_0^*$ with the $K^*$ off-shell scalar projection from the 
propagator in Eq.~(\ref{eqnprop}) shifts the $K_0^*$ resonance peak
to slightly smaller $Q^2$ values.
This effect appears to be probably too small to be established experimentally.
Another effect, however, of the scalar part of the $K^*$ off-shell
propagator is the dip close to the $K^*$ resonance $Q^2 = 0.8\,\mbox{GeV}^2$.
An experimental
confirmation of this dip would support the use of 
Eq.~(\ref{eqnprop}) for the $K^*$ propagator.
Of course,  $W_{SA}$ is very small in this mass region, and so only 
future high statistics experiment (b  or  tau-charm factories) could be
sensitive enough to study this effect.
As far as $W_{SF}$ is concerned, there is some sensitivity to the choice
of the off-shell propagator (compare the solid and the dotted-dashed 
curve). 
However, there is no qualitative difference, and in view of the uncertainty
in the precise value of $c_S$, it does not seem to be  possible 
to decide between
the two forms of the $K^*$ propagator from $W_{SF}$.

In Fig.~\ref{fig3}, we display results for the
three structure functions $W_B$, $W_{SA}$ and $W_{SF}$, normalized 
to $W_{tot}$.
We give results both with 
(Fig.~3a,b) and without (Fig.~3c,d)
the  $K_0^*$
and with (a,c) and without (b,d) the ${K^*}'$ contributions. 
Comparing Figs.~3a,b and Fig.~3c,d
we find that the scalar resonance leads to a strong enhancement of
$W_{SA}/W_{tot}$ above $Q^2 = 1.5\,\mbox{GeV}^2$ as
expected from Fig.~2a. 
The large negative value for the ratio of
$W_{SF}/W_{tot}$ in the range $Q^2=1-2$ GeV$^2$ 
is also an effect of the $K_0^*$ scalar resonance.

Comparing the results in (a,c) and (b,d), we
find that the value of $\beta_{K^*}$ does not change the overall
shapes of the $W_i/W_{tot}$ very much. 
Note that the nonvanishing contribution to
$W_{SA}/W_{tot}$ and $W_{SF}/W_{tot}$ in Fig.~3c,d is entirely due to
the interference of the scalar contribution from the off-shell $K^*$ with
the vector part.
The effect of the off-shell $K^*$ 
in the results shown in Fig.~3a,b is small.

Finally, we present the $Q^2$ distribution for the ratio of
$W_{SG}/W_{tot}$ in Fig.~4. 
Remember that $W_{SG}$ measures the 
imaginary part of the form factors ($FF_S^*$)
and requires nontrivial phases of the amplitudes.
These strong interaction phases are essential
for an observation of possible CP violation effects
in the difference of $W_{SG}$ for the $\tau^+$
and the $\tau^-$ decay \cite{argonne}.
Note however, that  $W_{SG}$ can only
be measured in experiments where the $\tau$ direction in the
laboratory frame can be determined as explained in Sec.~III.
{\vspace*{0.5cm} }

\section{Conclusions}

We have presented a meson dominance model for the the decay 
$\tau\to K\pi\nu_\tau$, which includes the vector resonances
$K^*(892)$, $K^*(1410)$ and the scalar $K_0^*(1430)$.
We have estimated the size of the $K_0^*$ contribution by matching
our model to the one loop prediction of chiral perturbation theory,
which determines the amplitude at very small momentum transfers.

Based on our model, we have calculated the vector and the scalar
form factors, which determine the hadronic matrix element.
The scalar form factor receives contributions both from the $K_0^*$
and from the off-shell spin-0 projection of the $K^*$.

We have presented predictions for the kaon-pion invariant mass
spectrum.
Although some indication for or against a sizable $K_0^*$ contribution
can be found in this spectrum, a clear disentanglement of the 
$K^*(1410)$ and the $K_0^*(1430)$ contributions
is possible only by analyzing the hadronic structure functions.
Both $W_{SA}$, the purely scalar structure function, 
and $W_{SF}$, which
measures  vector-scalar interference,
are interesting in this respect. 
A measurement of $W_{SA}(Q^2)$ allows for a clear establishment
of the $K_0^*$ contribution, while $W_{SF}(Q^2)$ yields additional
information about the scalar part, in particular its relative sign.

We also studied the issue of the off-shell $K^*$ propagator. 
We used a propagator for  massive vector mesons with
a spin-0 component in the off-shell region, rather than a strictly 
transverse propagator.
Matching of the vector meson dominance model to chiral
perturbation theory only works if this scalar component is included
in the vector meson propagator.
Experimentally, this scalar component could best be studied by measuring
$W_{SA}(Q^2)$ close to the $K^*(892)$ mass.
An explicit experimental confirmation, however,
is difficult and will require very high statistics.

\section*{Acknowledgements}
We would like to thank J.\ Smith for bringing this issue to
our attention and for helpful discussions. 
Interesting discussions and comments by J.H.~K\"uhn are
gratefully acknowledged.
The work of M.F.  was supported by the National Science
Foundation (Grant PHY-9218167) and by the Deutsche 
Forschungsgemeinschaft.
The work of E.~M. was supported in part  
by DFG Contract Ku 502/5-1.
%



\begin{thebibliography}{99}
\bibitem{personal}
There is some experimental evidence at CLEO of events at large 
$K^0 \pi^-$ mass beyond those expected
from $K^*(890)$.  How much of this is due to $K^0 K^-$
background is not yet clear, however
(J. Smith, personal communication).
\end{thebibliography}
\end{document}